\documentstyle[epsfig,amssymb]{mn}

\newif\ifAMStwofonts

\ifoldfss
  \ifCUPmtlplainloaded \else
    \NewTextAlphabet{textbfit} {cmbxti10} {}
    \NewTextAlphabet{textbfss} {cmssbx10} {}
    \NewMathAlphabet{mathbfit} {cmbxti10} {} 
    \NewMathAlphabet{mathbfss} {cmssbx10} {} 
  \fi
  \ifAMStwofonts
    \ifCUPmtlplainloaded \else
      \NewSymbolFont{upmath} {eurm10}
      \NewSymbolFont{AMSa} {msam10}
      \NewMathSymbol{\upi}     {0}{upmath}{19}
      \NewMathSymbol{\umu}     {0}{upmath}{16}
      \NewMathSymbol{\upartial}{0}{upmath}{40}
      \NewMathSymbol{\leqslant}{3}{AMSa}{36}
      \NewMathSymbol{\geqslant}{3}{AMSa}{3E}

    \fi
  \fi
\fi 

\ifnfssone
  \newmathalphabet{\mathit}
  \addtoversion{normal}{\mathit}{cmr}{m}{it}
  \addtoversion{bold}{\mathit}{cmr}{bx}{it}
  \newmathalphabet{\mathbfit} 
  \addtoversion{normal}{\mathbfit}{cmr}{bx}{it}
  \addtoversion{bold}{\mathbfit}{cmr}{bx}{it}
  \newmathalphabet{\mathbfss} 
  \addtoversion{normal}{\mathbfss}{cmss}{bx}{n}
  \addtoversion{bold}{\mathbfss}{cmss}{bx}{n}
  \ifAMStwofonts
    \ifCUPmtlplainloaded \else
      \UseAMStwoboldmath
      \makeatletter
      \new@mathgroup\upmath@group
      \define@mathgroup\mv@normal\upmath@group{eur}{m}{n}
      \define@mathgroup\mv@bold\upmath@group{eur}{b}{n}
      \edef\UPM{\hexnumber\upmath@group}
      \new@mathgroup\amsa@group
      \define@mathgroup\mv@normal\amsa@group{msa}{m}{n}
      \define@mathgroup\mv@bold\amsa@group{msa}{m}{n}
      \edef\AMSa{\hexnumber\amsa@group}
      \makeatother
      \mathchardef\upi="0\UPM19
      \mathchardef\umu="0\UPM16
      \mathchardef\upartial="0\UPM40
      \mathchardef\leqslant="3\AMSa36
      \mathchardef\geqslant="3\AMSa3E
    \fi
  \fi
\fi 

\ifnfsstwo
  \DeclareMathAlphabet{\mathbfit}{OT1}{cmr}{bx}{it}
  \SetMathAlphabet\mathbfit{bold}{OT1}{cmr}{bx}{it}
  \DeclareMathAlphabet{\mathbfss}{OT1}{cmss}{bx}{n}
  \SetMathAlphabet\mathbfss{bold}{OT1}{cmss}{bx}{n}
  \ifAMStwofonts
    \ifCUPmtlplainloaded \else
      \DeclareSymbolFont{UPM}{U}{eur}{m}{n}
      \SetSymbolFont{UPM}{bold}{U}{eur}{b}{n}
      \DeclareSymbolFont{AMSa}{U}{msa}{m}{n}
      \DeclareMathSymbol{\upi}{0}{UPM}{"19}
      \DeclareMathSymbol{\umu}{0}{UPM}{"16}
      \DeclareMathSymbol{\upartial}{0}{UPM}{"40}
      \DeclareMathSymbol{\leqslant}{3}{AMSa}{"36}
      \DeclareMathSymbol{\geqslant}{3}{AMSa}{"3E}
    \fi
  \fi
\fi 

\ifCUPmtlplainloaded \else
  \ifAMStwofonts \else 
    \def\upi{\pi}
    \def\umu{\mu}
    \def\upartial{\partial}
  \fi
\fi

\title[The optical counterpart of SAX J1808.4--3658] 
  {The optical counterpart of SAX J1808.4--3658, the transient bursting 
   millisecond X-ray pulsar}
\author[A. B. Giles et al.]
       {A. B. Giles, K. M. Hill, J. G. Greenhill\\
      School of Mathematics and Physics, University of Tasmania, 
      GPO Box 252-21, Hobart,
      Tasmania 7001, Australia}

\date{Accepted 1998 Nov 5.
      Received 1998 Oct 12;
      in original form 1998 July 27}

\pagerange{\pageref{firstpage}--\pageref{lastpage}}
\pubyear{1998}

\begin{document}

\maketitle

\label{firstpage}

\begin{abstract}
A set of CCD images have been obtained during the decline of the X-ray 
transient SAX J1808.4--3658 during April--June 1998. The optical
counterpart has been confirmed by several pieces of evidence. The optical 
flux shows a modulation on several nights which is consistent 
with the established X-ray binary orbit period of 2 hours. This optical 
variability is roughly in antiphase with the weak X-ray modulation. 
The source mean magnitude of {\it V\/}=16.7 on April 18 declined 
rapidly after April 22. From May 2 onwards the magnitude was more constant 
at around {\it V\/}=18.45 but by June 27 was below our sensitivity limit. 
The optical decline precedes the rapid second phase of the X-ray decrease 
by $3\pm1$ days. The source has been identified on a 1974 UK Schmidt plate 
at an estimated magnitude of $\sim$20. The nature of the optical companion 
is discussed.
\end{abstract}


\section{Introduction}
The X-ray transient SAX J1808.4--3658 was first detected by the Wide Field 
Camera (WFC) on the {\it BeppoSAX\/} X-ray satellite in September 1996 (in 't 
Zand et al. 1998). It was detectable for a period of $\sim$20 days but had 
not been visible during an earlier long exposure of the same region in August 
1996. Recently the Proportional Counter Array (PCA) experiment on the Rossi 
X-ray Timing Explorer {\it (RXTE)} satellite detected a transient 
(XTE J1808--369) in the same general location during a routine scan between 
targets. A series of Target Of Opportunity (TOO) observations with 
{\it RXTE\/} refined the error box and indicated it was probably a repeat 
occurrence of the earlier {\it SAX\/} transient outburst (Marshall 1998). 
The PCA data also revealed a high frequency signal at $\sim$400 Hz with 
a modulation of 4.3 per cent rms in the 2--60 keV band (Wijnands \& van der 
Klis 1998a,b). This is the first X-ray source to show coherent 
millisecond periodicity in its persistent emission. Further analysis 
of the PCA data showed this signal to be modulated with a sinusoidal 
doppler shift about a mean frequency of 400.9752106 Hz indicating a binary 
period of 7249.119 seconds (Chakrabarty \& Morgan 1998a,b). The X-ray flux 
has a weak modulation of 2 per cent with a broad minimum when the neutron 
star is behind the companion. The mass estimates for the companion suggest 
a very low value of probably $<$0.1 M$\sun$ (Chakrabarty \& Morgan 1998b). 
The X-ray flux from SAX J1808.4--3658 peaked close to April 11, just after 
the commencement of the {\it RXTE\/} series of observations. A relatively 
rapid X-ray decline after April 25 has been reported by Gilfanov, 
Revnivtsev \& Sunyaev (1998a) and Gilfanov et al. (1998b). 
At higher energies the High Energy X-ray Timing Experiment (HEXTE) on 
{\it RXTE\/} found the source to be one of the hardest known X-ray pulsars 
with a power law index of $2.02\pm0.05$ and a spectrum extending to at 
least 120 keV (Heindl, Marsden \& Blanco 1998, Heindl \& Smith 1998).

Roche et al. (1998) reported a probable optical counterpart with a magnitude 
of {\it V\/}=16.6 that was not on an earlier UK Schmidt Digitised Sky Survey 
plate dating from 1987. We had also commenced optical photometry of the X-ray 
error box and were able to support the proposed identification since the 
candidate appeared to show a $\sim$2 hour {\it V\/} band modulation of 
$\sim$0.12 magnitudes peak to peak (Giles, Hill \& Greenhill 1998). The 
{\it V\/} band intensity decreased by $\sim$0.1 magnitudes over the 
4 day period from April 18--22 further supporting the identification (Giles 
et al. 1998). The orbital doppler solution derived by Chakrabarty \& Morgan 
(1998a,b) is of sufficient precision to define a relatively small error box 
which also includes the candidate star. The assertion that the companion was 
a low mass star was supported by optical spectra obtained by Filippenko \& 
Leonard (1998). Their spectra revealed absorption lines throughout the 
spectrum which are characteristic of mid to late type stars. They also 
reported a possible broad $H\alpha$ emission line. A compilation of 
these other optical studies can be found in Chakrabarty et al. (1998c).

\section{Observations}
All the observations described in this paper were made using the Mt. Canopus 
1-m telescope at the University of Tasmania observatory. This telescope 
has been making CCD observations of micro lensing events for several years 
as part of the PLANET consortium. A new CCD camera has been installed  
and the data obtained on 1998 April 18 are "first light" observations. The SITe 
CCD chip is a thinned back illuminated device with a pixel size of 24 x 24 
microns. The chip has 512 x 512 pixels with an image scale of $0.42\arcsec$ 
pixel$^{-1}$. Standard {\it V\/} \& {\it I\/} filters (Bessell 1990) 
were used for the observations reported here. The CCD chip 
is driven by a Leach SDSU controller running under the CICADA software 
suite developed by the Mt. Stromlo and Siding Spring Observatories. 
The individual CCD images have been reduced using MIDAS software running 
under LINUX on a PC. The data values in Table 1 and Fig. 1 were derived 
using the crowded field photometry program DoPHOT to establish differential 
magnitudes between the candidate star and local secondary standards. 
Our best CCD frames were used to build a "fixed position" template for 
DoPHOT to fit an analytic point spread function to the stellar images. 
This method has been found to be superior when dealing with poor or 
variable seeing.   

We have followed the object from the date on which a small X-ray error box 
was announced (April 15) by Marshall et al. (1998) until late June. 
A journal of the observations is given in Table 1. The seeing on April 22 
(HJD 2450926) was poor (3$\arcsec$- 4$\arcsec$) and the observations on 
April 27 (HJD 2450931) were affected at times by thin cloud. When the 
source became fainter various program constraints on telescope time only 
allowed occasional exposures at a few wavelengths. Since the proposed 
optical candidate was not visible at an earlier epoch (Roche et al. 1998) 
we expected the star to fade from view and this eventually proved to be 
the case. The complete data set from April 18 to June 27 forms the subject 
of this paper.

\begin{table}
 \caption{A journal of the source observations.}
 \label{symbols}
 \begin{tabular}{@{}lcccccc}
        HJD  &  {\it V\/} mag.  &        HJD  &  {\it I\/} mag.  \\
  922.20473  &  $16.72$         &  922.19330  &  $16.11\pm0.05$  \\
  926.22798  &  $16.77$         &  931.24646  &  $17.36\pm0.13$  \\
  931.16728  &  $17.81$         &  933.25167  &  $17.65\pm0.08$  \\
  933.24373  &  $18.14\pm0.09$  &  933.26705  &  $17.69\pm0.10$  \\
  933.26326  &  $18.20\pm0.09$  &  936.20005  &  $17.83\pm0.10$  \\
  936.20443  &  $18.58\pm0.11$  &  938.19292  &  $17.83\pm0.13$  \\
  936.20814  &  $18.58\pm0.10$  &  939.16966  &  $18.12\pm0.15$  \\
  938.17917  &  $18.38\pm0.10$  &  950.17325  &  $17.80\pm0.15$  \\
  938.18353  &  $18.35\pm0.09$  &  950.99307  &  $17.86\pm0.08$  \\
  939.17961  &  $18.51\pm0.14$  &  951.00449  &  $17.83\pm0.08$  \\
  939.27461  &  $18.49\pm0.11$  &             &                  \\
  950.17826  &  $18.36\pm0.14$  &             &                  \\
  951.01020  &  $18.44\pm0.05$  &             &                  \\
  992.21600  &  $>20.5$         &             &                  \\
 \end{tabular}
 \medskip

The above HJD -2450000 times are the mid points of the typically 
300 second integration intervals. The first 3 values in the {\it V\/} 
magnitude column are averages for each of the 3 data sets shown in Fig. 1.
\end{table}

\begin{figure}
\epsfig{file=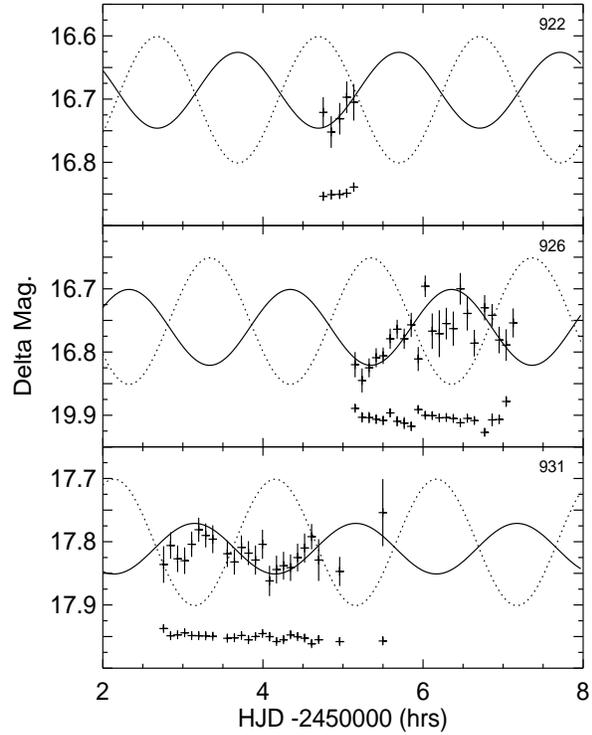, height=10.5cm, width=8.5cm }
 \caption{The three sections of light curves obtained from sequential 
{\it V\/} band CCD images are shown here in the three panels. The solid 
curve is the best estimate to the data using the X-ray binary period as 
a fixed parameter. The dotted trace is the X-ray ephemeris from 
Chakrabarty \& Morgan (1998b). The lower points in each panel indicate 
the constancy of local standard star 1 (see text).}
\end{figure}

\begin{figure}
 \epsfig{file=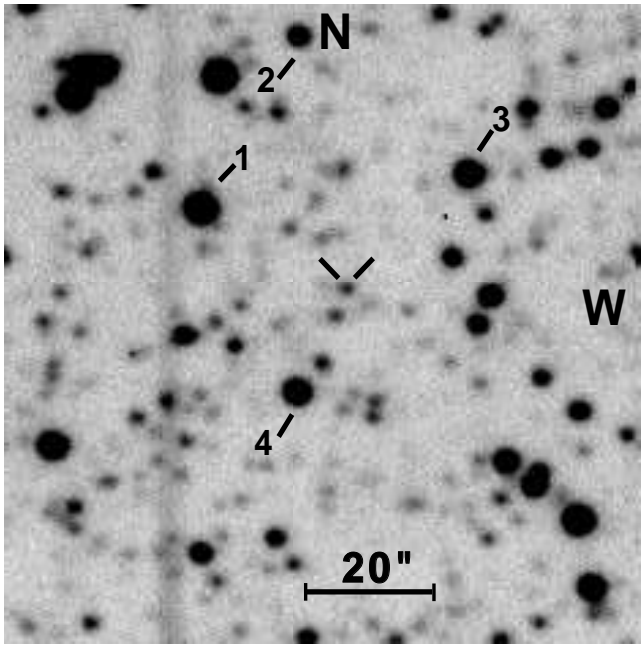, height=8.5cm, width=8.8cm }
 \caption{A finder chart for SAX J1808.4--3658. This composite {\it V\/} band 
image is from April 27 when the source was at {\it V\/}=17.8. The close 
neighbour mentioned by Roche et al. (1998), just 5$\arcsec$ to the SSE, 
is similar in brightness. The vertical stripe is from a bright star in the 
CCD frame which is outside the region shown here. The four secondary 
standards used are marked with the numbers 1-4.}
\end{figure}

\subsection{Temporal variability}
The data were calibrated by observations of a sequence of standard stars 
in Area Number 107 of Plate 57 (Landolt 1992). These were used to derive the 
magnitudes of four local secondary standards that were near to 
SAX J1808.4--3658 and within the CCD frame. These local standards are marked 
on the finder chart in Fig. 2. Our derived {\it V\/} magnitudes for stars 
1-4 are 14.57, 16.13, 15.21 \& 15.30 all $\pm$ 0.03. The corresponding 
four {\it I\/} magnitudes are 13.06, 15.20, 14.32 \& 13.70 all $\pm$ 0.025. 
The magnitudes for SAX J1808.4--3658, shown in Table 1 and plotted in the 
upper section of each panel in Fig. 1, were obtained using differential 
magnitudes relative to local standard star number 1. We present first 
the overall light curve decay and then the detailed light curves available 
for the first three nights. We also make some comments on the longer 
term variability. 

\subsubsection{The optical and X-ray light curves}
We have used the {\it V\/} \& {\it I\/} band data in Table 1 to produce 
the {\it V\/} \& {\it V--I\/} light curves shown in Fig. 3. The two 
intensities decreased in step with each other with no convincing evidence 
for colour change. The average uncorrected for reddening {\it V--I\/} 
value is $\sim$0.57 magnitudes The optical decline appears to have occurred 
mainly between April 22 and May 2. After this time the source remained 
roughly constant at around {\it V\/}=18.45 and {\it I\/}=17.86 for 
several weeks. On June 27 the source was below our sensitivity limit 
implying a {\it V\/} magnitude of less than $\sim$20.5. This last 
observation is not included in Fig. 3. 

The {\it RXTE\/} satellite made a series of 21 public TOO observations 
of SAX J1808.4--3658, the first occurring on April 11 and the last on May 6. 
In Fig. 3 we also plot the mean X-ray fluxes reported by Gilfanov 
et al. (1998b) and the two lines they obtain (see their Fig. 3) by 
treating the X-ray intensity profile as having two seperate linear decay 
regions. Our optical measurements commenced near the end of their first 
decay section since the peak X-ray intensity occurred around April 11 
(HJD 2450915.30). There is no X-ray intensity information to relate 
to the observed optical plateau after the last {\it RXTE\/} pointing 
on May 6 (HJD 2450939.95). In Fig. 3 the optical flux decline appears to 
precede the steep second part of the X-ray decline by $3\pm1$ days.

\begin{figure}
\epsfig{file=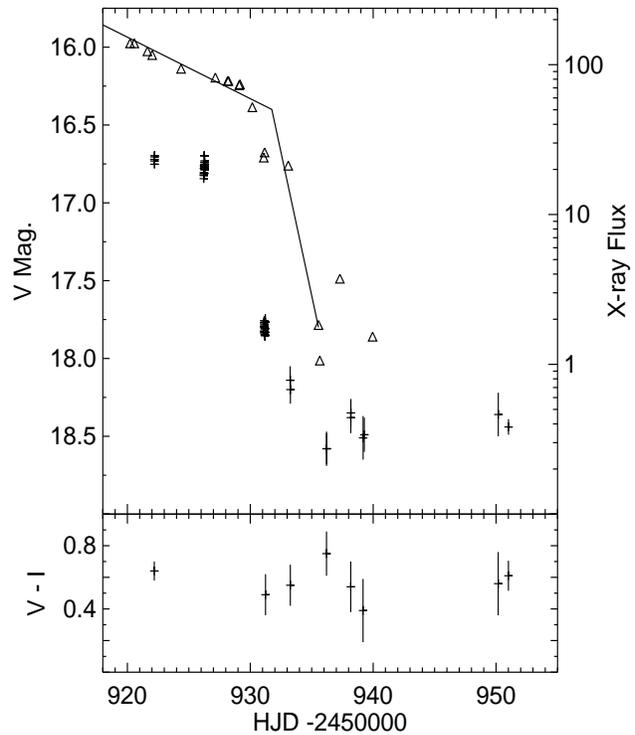, height=10.5cm, width=8.5cm}
  \caption{The {\it V\/} (+) \& X-ray ($\triangle$) decay light 
curves of SAX J1808.4--3658 during April and May 1998. The lines mark 
the X-ray decline as shown in Fig.1 of Gilfanov et al. (1998b). 
The {\it RXTE\/} PCA data covers the energy range 3-25 keV and the flux 
is in units of $10^{-11} erg$ $sec^{-1} cm^{2}$.}
\end{figure}

\subsubsection{The 2 hour binary modulation}
We have plotted the light curves shown in Fig. 1 for the three nights of April 
18, 22 \& 27 when data frames were obtained in continuous sequential fashion. 
The times have all been converted to Heliocentric Julian Date (HJD). 
Since our observations show the source decaying in intensity and its 
amplitude diminishing it is problematic how best to deal with the three short 
data sets. A sine type modulation had been evident in the April 22 data 
with a peak to peak magnitude of $0.12\pm0.02$ (Giles et al. 1998). 
This night however had degrading seeing conditions and a re-analysis of our 
entire data set, using a DoPHOT template which includes nearby blended 
stars, has substantially degraded the evidence for this sinusoidal modulation. 
In principle the use of a template minimises the errors in magnitude 
measurement caused by star blending. The seeing was so poor during the 
latter half of the April 22 observations that three neighbouring stars 
were partially blended with SAX J1808.4--3658. Two of these are very 
faint and just visible in Fig. 2 to the East and Southwest of 
SAX J1808.4--3658. The third neighbour is the star noted by Roche 
et al. (1998) $\sim$5$\arcsec$ to the South-Southeast and on April 22 
SAX J1808.4--3658 was partially blended with this star but was more than a 
magnitude brighter. Fig. 2 is an image from April 27 when the seeing was 
much better than on April 22 and SAX J1808.4--3658 had faded by about a 
magnitude. An exhaustive analysis of the data forces the 
conclusion that the much clearer modulation noted by Giles et al. (1998) 
was caused by a fortuitous combination of experimental errors. Nevertheless 
the data show significant variability and are not inconsistent with the 
7249.119 second sinusoidal modulation in X-rays seen by Chakrabarty \& 
Morgan (1998b). We assume therefore that this X-ray binary period also 
applies to the optical data. The X-ray radial velocity ephemeris of 
Chakrabarty \& Morgan (1998b) is represented by the dotted line in Fig. 1. 
with the amplitude units being arbitrary. We assume a 
similar modulation in the {\it V\/} band with peak to peak amplitude 0.12 
magnitudes in antiphase with the X-ray modulatiion. This is consistent with 
the standard model for Low Mass X-ray Binaries (LMXBs) (van Paradijs 1983, 
van Paradijs \& McClintock 1995) in which optical minimum occurs when the 
neutron star is at superior conjunction and the X-ray heated side of the 
companion is obscured.

The {\it V\/} band data and assumed underlying modulation are plotted in the 
middle panel of Fig. 1. In the top panel the assumed modulation has been 
projected back to April 18 (HJD 2450922) with the same phase and amplitude 
but with a 0.05 magnitude offset to match the higher observed intensity at 
that time. The data sequence for April 18 is too short to define the 
modulation but we note that the trend is correct for the phase of the 
assumed modulation relative to the data on April 22. On April 27 
(HJD 2450931) the source was $\sim$1 magnitude fainter in {\it V\/} 
relative to April 22 and the modulation was less with a peak to peak 
amplitude of $\sim$0.08 magnitudes. The assumed modulation in the bottom 
panel of Fig 2 has been adjusted accordingly. Again the  {\it V\/} band 
modulation is consistent with our assumptions. 

We also plot in each panel of Fig. 1 the differences between local 
standard 1 and the sum of local standards 2, 3 \& 4. These points are 
offset to an arbitary magnitude since they are only intended to indicate 
the relative photometric precision on each of the three nights.

\subsubsection{Long term history}
We have examined a UK SERC J Schmidt sky survey film copy of this field 
(Plate number J 663, Field name J 394, date June 20 1974) and find a 
definite star image that corresponds to our source position. Estimating a 
magnitude to compare with those in Table 1 is difficult. We have used 
a digital camera to image the film copy and compared SAX J1808.4--3658 with 
the 3 stars in Fig. 2 that are visible within $8\arcsec$. We estimate a 
magnitude of $\sim$20, as the image is well above the plate limit, which is 
suggestive of an outburst in progress during May to June 1974. Clearly the 
source has an interesting long term intensity history having had the 
two known X-ray transient events of September 1996 and April 1998. Since 
SAX J1808.4--3658 is present on at least one Schmidt plate from 1974, is 
not present on the 1987 digitised sky survey image, and has again faded 
below detectability, a detailed search of all the available Schmidt plates 
of this region would be a worthwhile  exercise. The UK Schmidt catalogue 
lists 9 plates of this field and there are several additional ESO Schmidt 
plates.

\subsection{Source position}
As part of the photometry process the DoPHOT analysis software fits 2 
dimensional power law profiles to the target and specified reference star 
positions. This automatically provides accurate values for the relative 
positions of each star centre on the CCD frame in pixel coordinates. 
The $4\arcmin$ x $4\arcmin$ CCD field contains 6 stars from the 
Hubble Guide Star Catalogue (GSC) (Lasker et al. 1990). We have used the 
CCD pixel coordinates of these stars and SAX J1808.4--3658 on the best 
quality {\it V\/} band image to derive a source position of 
R.A. 18h 08m $27\fs54 \pm 0\fs015$ 
Dec. $-36\degr 58\arcmin 44\farcs3 \pm 0\farcs2$ (equinox J2000.0). 
This is $\sim$2$\farcs4$ from the initial position estimate of 
Roche et al. (1998). We note that one of our 6 reference stars (GSC 7403-0236) 
was deleted from our list as it produced an inconsistent result in the fitting 
procedure. This star is the closest of the 6 to the source position but 
appears to have an accumulated proper motion of 3$\farcs1$ to the South East 
over the $\sim$23 year interval. The CCD frame differential values are good 
to $\pm0\farcs2$ but the absolute position is dominated by the errors in 
the GSC which are up to $\sim\pm0\farcs8$. In Fig. 2 we provide a finder 
chart for SAX J1808.4--3658 constructed from a good seeing {\it V\/} band 
CCD image. The fainter star noted by Roche et al. (1998) with {\it V\/} 
$\sim$18 is 5$\arcsec$ to the South-Southeast.

\section{Discussion}
At first sight SAX J1808.4--3658 looks like any other LMXB transient but 
it has the unique feature of having a coherent millisecond rotation frequency 
in its X-ray emission. The X-ray spectrum was also reported to be remarkably 
constant over two orders of intensity decline (Gilfanov et al. 1998a,b) 
though this has been questioned by Heindl and Smith (1998). The binary 
orbit optical modulation shown in Fig. 1 is reminiscent of a number of 
other LMXB's such as 4U 1636-53 (Smale \& Mukai 1988), 4U 1735-44 (Corbet 
et al. 1986), X1755-338 (Mason, Parmar \& White 1985) and XB 1254 -690 
(Motch et al. 1987). Particularly in the case of 4U 1636-53 \& 4U 1735-44 
a sinusoidal modulation is well shown at some times and poorly followed at 
others. The fragments of light curves in Fig. 1 are suggestive of the same 
behaviour. 
  
Traditionally there are three regions of a LMXB system which contribute 
to its optical variability due to X-ray heating. These are the accretion 
disc itself, a bright spot on the outer edge of the accretion disc due to 
inflowing material and the hemisphere of the companion facing the neutron 
star which is not shadowed by the accretion 
disc. As remarked earlier the phasing of the observed optical modulation 
and the well defined X-ray orbital period are consistent with the established 
model for X-ray reprocessing on the facing hemisphere of the companion star. 
In most LMXBs the reprocessed X-ray optical flux dominates the the optical 
light from the rest of the system (van Paradijs 1983, van Paradijs \& 
McClintock 1995), particularly in the outburst phase, and the 
companion itself may only be evident at a very 
faint level when the system is in quiescence. Various approximately 
linear relationships between the orbit period, outburst amplitude, absolute 
magnitude and distance have been derived for LMXB transients by 
Shahbaz \& Kuulkers (1998). Although it is unclear if these relationships 
can be applied to such a low mass and short period system we obtain 
a pre-outburst magnitude of {\it V\/}=28.75 for SAX J1808.4--3658. 
These relationships also give a quiescence absolute magnitude of 
M(2) = 13.3 for the companion and M(disc) = 4.65. The companion peak 
apparent magnitude of {\it V\/} $\la$16.7 implies a distance of $<$2.5 kpc. 
Given the uncertainties in the Shahbaz \& Kuulkers (1998) relationships, 
this distance is not inconsistent with that deduced from the observation 
of two Type I X-ray bursts during the 1996 outburst which indicate 
a distance of 4 kpc (in 't Zand et al. 1998).

The optical decline appears to precede the X-ray decline by $3\pm1$ days. 
This delay could be interpreted as the time taken to clear out the accretion 
disc round the neutron star after a sharp drop in the mass flow rate, by 
roche lobe overflow or stellar wind, from the companion. The delay then 
represents a flow transit time from the outer to the inner edge of the disc. 
This explanation also suggests that the disc hot spot might decrease 
in intensity as the general accretion rate diminishes following 
the main transient event. Our observations show that both the optical 
intensity and its modulation amplitude at the binary period decrease. 
The sudden and rapid X-ray decline then follows since the disc is now 
mostly depleted, with little remaining material available to spiral in 
towards the neutron star, and a low external accretion rate to replenish it. 
The sharp X-ray decline is conventionally expected for LMXB transients, and 
can be explained in terms of the closure of a centrifugal barrier at 
low accretion rates as the magnetosphere of a neutron star reaches 
its corotation radius (Campana et al. 1998). In the case of SAX J1808.4--3658 
we then see a plateau in the optical flux lasting $\ga$2 weeks but there are 
no simultaneous X-ray observations. A final optical decline over the few 
weeks after this plateau is not constrained by our observations.

\section{Conclusions}
Our observations have established beyond doubt that the candidate proposed 
by Roche et al (1998) as the optical counterpart for the X-ray transient 
SAX J1808.4--3658 is correct. This has been demonstrated by the following 
key points as the proposed object:

\begin{tabbing}
xx\= \kill
\> (i) \ \ is within the various X-ray error boxes\\
\> (ii) \ faded by more than 4 magnitudes\\
\> (iii) showed a substantial part of its optical fading a\\
\> \ \ \ \ \ \ few days prior to the large X-ray decline\\
\> (iv) was modulated in intensity in a manner roughly\\
\> \ \ \ \ \ \ consistent with the 2 hr X-ray binary period.
\end{tabbing}
Since this source was seen in an earlier transient outburst during September 
1996 (in 't Zand et al. 1998) it has most probably burst before and will do 
so again. The optical flux of SAX J1808.4--3658 is expected to brighten on 
all these occasions so further observations, particularly spectroscopy while 
in its brightest state, are highly desirable,

\section*{Acknowledgments}
We thank K. Bolton, D. Phythian and B. Wilson of the Physics Department 
technical support group for their invaluable work on the CCD camera system. 
P. Cieslik kindly obtained some measurements between PLANET observations. 
We also thank T. Strohmayer of the {\it RXTE\/} PCA team for timely 
information on the occurrence of the transient.

\bsp 

\label{lastpage}

\end{document}